\documentclass[pre,11pt,nofootinbib]{revtex4}
\usepackage{amsmath}
\usepackage{graphicx}
\newcommand{\smfrac}[2]{\mbox{\small $#1 \over #2$}}
\newcommand{\chiSG}{\chi_{_{SG}}}
\newcommand{\av}{_\text{av}}
\begin{document}
\title{Notes on the replica symmetric solution of the classical and quantum SK model, including
the matrix of second derivatives and the spin glass susceptibility.}
\author{A. P. Young}
\affiliation{University of California Santa Cruz, CA 95064, USA}
\begin{abstract}
A review of the replica symmetric solution of the classical and quantum, infinite-range,
Sherrington-Kirkpatrick spin glass is presented.
\end{abstract}
\maketitle

\section{Introduction}
\label{sec:intro}
These notes assemble together many results on the replica symmetric (RS) solution
of the infinite-range Sherrington-Kirkpatrick~\cite{sherrington:75} (SK)
model. The quantum version, in which a transverse field is added, will be
discussed in detail, as well as the original classical version.
Little here is
original, and the bibliography indicates original sources. Some of the material
is taken from an old review article~\cite{binder:86}.

In fact, the (RS)
solution is unstable below the critical point, and the correct solution, which
is much more complicated, was found by Parisi~\cite{parisi:80,parisi:83},
several years after the
model was originally proposed. In a magnetic field, there is a line of
transitions, first found by de Almeida and Thouless~\cite{almeida:78},
in the temperature-field plane below which the RS solution is
unstable. This is known as the AT line. Almeida and Thouless obtained this
line by looking at the stability of the RS solution with respect to
fluctuations in the order parameters. Here, we shall discuss this stability
matrix, both in the quantum and classical cases. At the point where the RS solution goes
unstable, a response function called the spin glass susceptibility, $\chiSG$,
diverges. We shall compute $\chiSG$ for both the classical and quantum case.
Our expression for $\chiSG$ in the quantum case, seems to be new; probably the
\textit{only} new result in these notes. 

\section{The classical SK model}
\label{sec:class}

The Sherrington Kirkpatrick (SK)~\cite{sherrington:75}
model aims to provide a mean field solution
of the spin glass problem as the exact solution of an infinite range model.
The Hamiltonian is
\begin{equation}
\mathcal{H} = - \sum_{\langle i, j \rangle} J_{ij} S_i S_j\, ,
\end{equation}
where the $S_i$ are Ising spins, $s_i = \pm 1$, and the $J_{ij}$ are independent random
variables with the \textit{same} distribution for all pairs $i$ and $j$,
\begin{equation}
P(J_{ij}) =  {1 \over J}\, \sqrt{N \over 2 \pi}\, 
\exp\left(-N J_{ij}^2 / 2 J^2\right) \, ,
\end{equation}
so the mean is zero and the standard deviation is $J$. The spin glass
transition temperature is at
\begin{equation}
T_c = J.
\label{Tc}
\end{equation}

The model is solved by
the replica trick, see e.g.~\cite{edwards:75}, according to which one calculates
the average free energy, i.e.\ the average of the \textit{log}
of the partition function $Z$, from the average of the $n$-th power
of $Z$ in the limit $n \to 0$. One has
\begin{subequations}
\begin{align}
[Z^n]\av &= 
\prod_{\langle i , j \rangle} \left[ \int_{-\infty}^\infty P(J_{ij}) \, d J_{i, j} \right]  \, 
\sum_{\{S_i^\alpha=\pm 1\}}\,
\exp\left[\beta \sum_{\langle i, j\rangle} J_{ij} \sum_{\alpha = 1}^n
S_i^\alpha S_j^\alpha \right] , \\
&=
\sum_{\{S_i^\alpha=\pm 1\}}\, 
\exp\left[{(\beta J)^2 \over 2 N} \sum_{\langle i, j\rangle}
\sum_{\alpha, \beta= 1}^n
S_i^\alpha S_j^\alpha S_i^\beta S_j^\beta \right] \, ,
\label{Zn0}
\end{align}
\end{subequations}
where $[\cdots]\av$ denotes an average over the quenched bond disorder. 
Separating out the $\alpha = \beta$ terms, and dropping some $1/N$ corrections
gives
\begin{equation}
[Z^n]\av =  \exp\left[\smfrac{1}{4}(\beta J)^2 N n \right] \,
\sum_{\{S_i^\alpha=\pm 1\}}\,
\exp\left[{(\beta J)^2 \over 2 N} \sum_{\alpha < \beta} 
\left( \sum_i S_i^\alpha S_i^\beta\right)^2 \right] \, .
\label{Zn1}
\end{equation}
We decouple the square using the Hubbard-Stratonovich transformation for each
pair of indices $\alpha < \beta$,
\begin{equation}
e^{\lambda a^2 / 2}= \sqrt{\lambda\over 2 \pi} \,
\int_{-\infty}^\infty dx \, \exp\left[-\lambda {x^2 \over 2} + a \lambda x
\right] \, 
\label{HS}
\end{equation}
with
\begin{equation}
\lambda = N, \qquad a = {\beta J \over N} \, \sum_i S_i^\alpha S_i^\beta \, ,
\qquad x = (\beta J)\, q_{\alpha\beta} \,
\end{equation}
which gives
\begin{equation}
\begin{split}
[Z^n]\av =  \exp\left[\smfrac{1}{4}(\beta J)^2 N n \right] \,
\prod_{\alpha < \beta} & \left[\sqrt{N\over 2 \pi}\,
(\beta J) \int_{-\infty}^\infty 
d q_{\alpha\beta}\right] \times \\
& \exp\left[
-N {(\beta J)^2 \over 2} \sum_{\alpha < \beta} q_{\alpha\beta}^2 + N \ln
\mathrm{Tr}
\exp [-H] \, \right] \, ,
\end{split}
\label{Zn}
\end{equation}
where
\begin{equation}
H = -(\beta J)^2 \sum_{\alpha < \beta} q_{\alpha \beta} S^\alpha S^\beta \, ,
\label{H_1}
\end{equation}
and the trace is now over the $n$ Ising spins $S^\alpha$. The $N$ sites $i$
have been decoupled and so the trace over each
spin gives the same result, namely $ \mathrm{Tr} \exp (-H)$.
We can write Eqs.~\eqref{Zn} and \eqref{H_1} as 
\begin{equation}
\left[Z^n \right]\av = \prod_{\alpha < \beta} \left[\sqrt{N\over 2 \pi}\,
(\beta J) \int_{-\infty}^\infty d q_{\alpha\beta}\right] \,
\exp\left[ - N n \beta f(q) \right] \, ,
\label{Zn2}
\end{equation}
where
\begin{equation}
-\beta f(q) = \lim_{n\to0} \left[ {(\beta J)^2 \over 4} -
{(\beta J)^2\over 4 n} \sum_{(\alpha,\beta)} q^2_{\alpha\beta} + {1 \over n}
\log \mathrm{Tr} e^{-H} \right] \, ,
\label{fab}
\end{equation}
where the notation $(\alpha,\beta)$ means sum over all distinct replicas (so
each pair is counted \textit{twice}).
Because of the overall factor of $N$ in the exponent in Eq.~\eqref{Zn2}, we will evaluate the
integrals by the method of steepest descent. Neglecting subleading terms, the
answer is just the exponential in Eq.~\eqref{Zn2} with the $q_{\alpha\beta}$
evaluated at the saddle point, i.e.~the $q_{\alpha\beta}$ are given by a
self-consistent solution of
\begin{equation}
q_{\alpha\beta} = {\mathrm{Tr}\, S^\alpha S^\beta\, e^{-H} \over
\mathrm{Tr}\, e^{-H}} \qquad \left(= \langle S^\alpha S^\beta \rangle  \right) \, ,
\label{qab}
\end{equation}
where the average $\langle \cdots \rangle$ is with respect to the weight
$e^{-H}$
with the $q_{\alpha\beta}$ taking their saddle point values.


We look for the replica-symmetric solution where each of the $n(n-1)/2$ order
parameters $q_{\alpha\beta}$ takes the same value $q$. In this case
\begin{equation}
H =-\smfrac{1}{2} (\beta J)^2 q \sum_{(\alpha,\beta)} S^\alpha S^\beta 
= \smfrac{1}{2} (\beta J)^2 q \left[
\left(\sum_\alpha S^\alpha\right)^2 - n \right] \, ,
\label{L_RS}
\end{equation}
so
\begin{equation}
-\beta f = \lim_{n\to 0} \left[ {(\beta J)^2 \over 4}(1-q)^2  
+ {1 \over n} \ln \mathrm{Tr} \,
\exp\left[{(\beta J)^2 \over 2} q\,
\left(\sum_\alpha S^\alpha \right)^2
\right]\, \right] \, .
\label{eqA}
\end{equation}
We decouple the term quadratic in the spins by another Hubbard-Stratonovich
transformation, Eq.~\eqref{HS}, with $\lambda =1, a = \beta J q^{1/2}
\sum_\alpha S^\alpha, x = z$, i.e.
\begin{equation}
\exp\left[{(\beta J)^2 \over 2} q\,
\left(\sum_\alpha S^\alpha \right)^2 \right] = 
{1 \over \sqrt{2 \pi} } \int_{-\infty}^\infty dz \, e^{-z^2/2}
\, e^{\beta J q^{1/2}\, z\, \sum_\alpha S^\alpha}.
\label{expH}
\end{equation}
Hence
\begin{subequations}
\label{zSa_all}
\begin{align}
\label{zSa}
\mathrm{Tr} \,
\exp\left[{(\beta J)^2 \over 2} q\,
\left(\sum_\alpha S^\alpha \right)^2 \right] &= 
{1 \over \sqrt{2 \pi} } \int_{-\infty}^\infty dz \, e^{-z^2/2}\, \mathrm{Tr}
\, e^{\beta J q^{1/2}\, z\, \sum_\alpha S^\alpha} \\
&= {1 \over \sqrt{2 \pi}} \int_{-\infty}^\infty dz \, e^{-z^2/2} \, \left(
2 \cosh \beta J q^{1/2} \, z\right)^n \\
&= 1 + n  {1 \over \sqrt{2 \pi}} \int_{-\infty}^\infty dz \, e^{-z^2/2} \,
\ln (2 \cosh \beta J q^{1/2}\,z) + O(n^2) \, ,
\end{align}
\end{subequations}
where the last line expands the result in powers of $n$. Substituting into
Eq.~\eqref{eqA} and taking the limit $n \to 0$ gives the free energy of the SK
model in the replica symmetric ansatz as
\begin{equation}
\boxed{
-\beta f = {(\beta J)^2 \over 4}(1-q)^2  + 
{1 \over \sqrt{2 \pi}} \int_{-\infty}^\infty dz \, e^{-z^2/2} \,
\ln (2 \cosh \beta J q^{1/2}\,z) \, .}
\end{equation}
The order parameter $q$ is obtained by finding an extremal
value of $f$. This gives
\begin{equation}
{(\beta J)^2 \over 2}\, (1 - q)
= {1 \over \sqrt{2 \pi}} \, {\beta J \over 2 q^{1/2}}\,
\int_{-\infty}^\infty dz \, e^{-z^2/2} \, z \, \tanh (\beta J q^{1/2}\,z) \, .
\end{equation}
Integrating by parts gives the final self-consistent equation for $q$.
\begin{equation}
\boxed{
q = {1 \over \sqrt{2 \pi}} \int_{-\infty}^\infty dz \, e^{-z^2/2} \,
\tanh^2(\beta J q^{1/2}\, z) \, .}
\label{q_sc}
\end{equation}
This can also be derived by noting that $q \ (=q_{\alpha\beta}) = 
\langle S^\alpha S^\beta\rangle$, where the average is over the weight $e^{-H}$,
see Eq.~\eqref{qab}. Using Eqs.~\eqref{L_RS} and \eqref{expH}, one readily
obtains Eq.~\eqref{q_sc}.

Next we consider fluctuations about the saddle point, i.e.
\begin{equation}
q_{\alpha\beta} = q + \delta q_{\alpha\beta} \, .
\end{equation}
The first derivative of $f$ with respect to $q_{\alpha\beta}$ is zero so we go
to the second derivative of $f$ in Eq.~\eqref{fab}, i.e.
\begin{equation}
\boxed{
\beta f[\{q\}] = \beta f[q^c] + \lim_{n \to 0} {1 \over n}\, {1 \over 2}
\sum_{\alpha<\beta,\gamma < \delta} A^{\alpha\beta,\gamma\delta} 
\delta q_{\alpha\beta}\, \delta q_{\gamma\delta} \, . }
\label{fq_quadratic}
\end{equation}
where, from Eq.~\eqref{fab},
\begin{align}
A^{\alpha\beta,\gamma\delta} & \equiv {\partial^2  (\beta f) \over \partial
q_{\alpha\beta} \partial q_{\gamma\delta} } \ \ (\times n) \, , \nonumber \\
&\boxed{= (\beta J)^2 \delta_{\alpha\beta,\gamma\delta} - (\beta J)^4
\left[
\langle S^\alpha S^\beta S^\gamma S^\delta\rangle - \langle S^\alpha
S^\beta\rangle \langle S^\gamma S^\delta\rangle
\right] \, ,}
\label{Aab}
\end{align}

Firstly consider $T > T_c$, where $H = 0$, so  $\langle S^\alpha S^\beta
\rangle = 0$ and hence only the $(\alpha,\beta) =
(\gamma,\delta)$ term contributes. Thus
\begin{equation}
A^{\alpha\beta,\gamma\delta} = \delta_{\alpha\beta,\gamma\delta} 
\,(\beta J)^2 \,\left(1 - (\beta J)^2 \right) \, .
\end{equation}
Now $A$ is a matrix of size $n(n-1)2$ so above $T_c$ all $n(n-1)/2$
eigenvalues are equal and given by
\begin{equation}
\boxed{
\lambda = (\beta J)^2 \left(1 - (\beta J)^2 \right) \quad(T > T_c) \, .
}
\end{equation}

Now we consider $T < T_c$. There are three types of term:
\begin{itemize}
\item $(\alpha\beta) (\alpha\beta)$. \\
Here we have $S^\alpha S^\beta S^\gamma
S^\delta = 1$, and $\langle S^\alpha S^\beta\rangle = \langle S^\gamma
S^\delta\rangle = q$.  Hence
\begin{equation}
(\beta J)^{-2} A^{\alpha\beta,\alpha\beta} = 1 - (\beta J)^2 (1 - q^2)
\qquad(= P, \ \text{say}) \, .
\label{abab}
\end{equation}
\item $(\alpha\beta) (\alpha\gamma)$ with $\beta \ne \gamma$. \\
Now $\langle S^\alpha S^\beta S^\alpha S^\gamma\rangle = \langle S^\beta
S^\gamma \rangle = q$. Also $\langle S^\alpha S^\beta \rangle = \langle
S^\gamma S^\delta \rangle = q$.  Hence
\begin{equation}
(\beta J)^{-2} A^{\alpha\beta,\alpha\gamma} = - (\beta J)^2 (q - q^2) \qquad
( = Q,\ \text{say}) \, .
\label{abac}
\end{equation}
\item $(\alpha\beta) (\gamma\delta)$ with all indices different.\\
As before $\langle S^\alpha S^\beta \rangle = \langle S^\gamma S^\delta \rangle
= q$. What about $\langle S^\alpha S^\beta S^\gamma S^\delta\rangle$? From
Eq.~\eqref{zSa} we see that the (unnormalized) probability distribution for the
$\{S^\alpha\}$ is 
\begin{equation*}
{1 \over \sqrt{2 \pi} } \int_{-\infty}^\infty dz \, e^{-z^2/2}
\, e^{\beta J q^{1/2}\, z\, \sum_\alpha S^\alpha} \, .
\end{equation*}
Hence
\begin{align}
\langle S^\alpha S^\beta S^\gamma S^\delta\rangle &=
\lim_{n \to 0} \left[ 
{1 \over \sqrt{2 \pi} } \int_{-\infty}^\infty dz \, e^{-z^2/2}
\, \mathrm{Tr}\,
S^\alpha S^\beta S^\gamma S^\delta\, e^{\beta J q^{1/2}\, z\, \sum_\alpha S^\alpha}
\over
{1 \over \sqrt{2 \pi} } \int_{-\infty}^\infty dz \, e^{-z^2/2}
\,\mathrm{Tr}\, e^{\beta J q^{1/2}\, z\, \sum_\alpha S^\alpha}
\right] \nonumber \\
&=
\lim_{n \to 0} \left[ 
{1 \over \sqrt{2 \pi} } \int_{-\infty}^\infty dz \, e^{-z^2/2} \,
\sinh^4 (\beta J q^{1/2} z)\, \cosh^{n-4} (\beta J q^{1/2} z)
\over
{1 \over \sqrt{2 \pi} } \int_{-\infty}^\infty dz \, e^{-z^2/2}
\cosh^n(\beta J q^{1/2} z)
\right] \nonumber \\
&= {1 \over \sqrt{2 \pi} } \int_{-\infty}^\infty dz \, e^{-z^2/2} \,
\tanh^4(\beta J q^{1/2} z) \, \qquad (= r\ \text{say}).
\label{abcd}
\end{align}
Hence, for all indices different, 
\begin{equation}
(\beta J)^{-2} A^{\alpha\beta,\gamma\delta} = - (\beta J)^2 (r - q^2) \qquad
( = R,\ \text{say}) \, .
\end{equation}
\end{itemize}

According to de Almeida and Thouless~\cite{almeida:78},
the important eigenvalue, the one which
goes negative, is the ``replicon'' mode,
$\lambda_r$, where 
\begin{equation}
\lambda_r = (\beta J)^2 \left[P - 2 Q  + R\right] .
\label{lar}
\end{equation}
Hence
\begin{align}
\lambda_r &=(\beta J)^2 \left[ 1 - (\beta J)^2(1 - q^2) + 2 (\beta J)^2 (q - q^2) - (\beta J)^2
(r - q^2) \right] \nonumber \\  
& = (\beta J)^2 \left[ 1 - (\beta J)^2(1 - 2 q + r)\right] \, ,
\label{lambda_r}\\
& \boxed{ =(\beta J)^2 \left\{ 1 - (\beta J)^2 \, {1 \over \sqrt{2\pi}}\int_{-\infty}^\infty
dz \, e^{-z^2/2} \,
\left[1 - \tanh^2(\beta J q^{1/2} z)
\right]^2 \right\} \, .}
\label{lambda_r_b}
\end{align}

Next we compute the spin glass susceptibility $\chiSG$ defined by
\begin{equation}
\chiSG = {1 \over T^2}{1 \over N} \sum_{i, j = 1}^N \left[ 
\left( \langle S_i S_j \rangle - \langle S_i \rangle \langle S_j \rangle \right)^2
\right]\av \, .
\label{chisg}
\end{equation}
Frequently the factor of $1/T^2$ is omitted because it varies smoothly near a
classical transition at finite $T$ but here we will eventually consider a
quantum transition at $T=0$ so we include it.

Considering the terms separately, it is standard to show, see e.g.~Binder
and Young~\cite{binder:86}, that they can be expressed as
\begin{subequations}
\label{chisg_terms}
\begin{align}
\left[ \langle S_i S_j\rangle^2 \right]\av &= \lim_{n\to 0}
{1 \over n(n-1)} \sum_{(\alpha, \beta)} \langle S_i^\alpha S_j^\alpha S_i^\beta
S_j^\beta\rangle,  \\
\left[ \langle S_i S_j\rangle \langle S_i\rangle \langle S_j\rangle \right]\av &= \lim_{n\to 0}
{1 \over n(n-1)(n-2)} \sum_{(\alpha, \beta, \gamma)} \langle S_i^\alpha S_j^\alpha S_i^\beta
S_j^\gamma\rangle,  \\
\left[ \langle S_i\rangle^2 \langle S_j\rangle^2 \right]\av &= \lim_{n\to 0}
{1 \over n(n-1)(n-2)(n-3)} \sum_{(\alpha, \beta, \gamma, \delta)}
\langle S_i^\alpha S_j^\beta S_i^\gamma S_j^\delta\rangle,  
\end{align}
\end{subequations}
where the averages on the RHS are with respect to the weight factor in
Eq.~\eqref{Zn0} and the notation $(\alpha, \beta)$ etc.~means all distinct
sets of replicas are to be summed over. Note that each thermal average on the
LHS of Eqs.~\eqref{chisg_terms} corresponds to a distinct replica on the RHS.

To calculate these averages we add a set of fictitious fields
$\Delta_{\alpha\beta}$ which couple to $\sum_i S^\alpha_i S^\beta_i$, i.e.
\begin{equation}
[Z^n]\av = 
\sum_{\{S_i^\alpha=\pm 1\}}\, 
\exp\left[{(\beta J)^2 \over 2 N} \sum_{\langle i, j\rangle}
\sum_{\alpha, \beta= 1}^n
S_i^\alpha S_j^\alpha S_i^\beta S_j^\beta  + 
\sum_{\alpha < \beta} \Delta_{\alpha\beta}\sum_i S^\alpha_i S^\beta_i \right] \, .
\end{equation}
Note that for $n \to 0$ there is no normalizing denominator so
\begin{subequations}
\label{derivs}
\begin{align}
\sum_i \langle S_i^\alpha S_i^\beta\rangle &= \lim_{n\to 0} {\partial \over \partial
\Delta_{\alpha\beta}}\left[ Z^n \right]\av, \\
\sum_{i,j} \langle S_i^\alpha S_i^\beta S_j^\gamma S_j^\delta \rangle &=
\lim_{n\to 0}
{\partial^2 \over \partial \Delta_{\alpha\beta} \Delta_{\gamma\delta}}
\left[ Z^n \right]\av, 
\end{align}
\end{subequations}
in which the replicas $\alpha,\beta,\gamma,\delta$ can take any values subject
to the restrictions $\alpha < \beta, \gamma < \delta$.
We note that $[Z^n]\av$ is still given by
Eq.~\eqref{Zn2}, with $\beta f(q)$ still given by \eqref{fab}, but now
\begin{equation}
H = -(\beta J)^2 \sum_{\alpha < \beta} \left[q_{\alpha \beta} -
(\beta J)^{-2} \Delta_{\alpha\beta}\right] S^\alpha S^\beta \, .
\end{equation}

We define shifted variables $u_{\alpha\beta}$ by
\begin{equation}
u_{\alpha\beta} = q_{\alpha\beta} - (\beta J)^{-2} \Delta_{\alpha\beta} \, ,
\end{equation}
so the $\Delta_{\alpha\beta}$ no longer appear in $H$ but rather in the
quadratic term in Eq.~\eqref{fab}, so
\begin{equation}
-\beta f(q) = \lim_{n\to 0} \left[ {(\beta J)^2 \over 4} -
{1\over 4 n} \sum_{(\alpha,\beta)} \left\{ (\beta J)^2 u^2_{\alpha\beta}
+ 2 \Delta_{\alpha\beta} u_{\alpha\beta} + (\beta J)^{-2}
\Delta^2_{\alpha\beta} \right\}  + {1 \over n}
\log \mathrm{Tr} e^{-H} \right] \, ,
\label{bf2}
\end{equation}
where now 
\begin{equation}
H = -(\beta J)^2 \sum_{\alpha < \beta} u_{\alpha \beta} 
S^\alpha S^\beta \, ,
\end{equation}
and the $u_{\alpha\beta}$ have to be integrated like the
$q_{\alpha\beta}$ in Eq.~\eqref{Zn2}. Performing the derivatives in
Eqs.~\eqref{derivs} we get
\begin{subequations}
\begin{align}
{1 \over N} \sum_i \langle S_i^\alpha S_i^\beta \rangle &= \lim_{n \to 0} 
\langle q_{\alpha\beta} \rangle , \\
{1 \over N} \sum_{i, j} \langle S_i^\alpha S_j^\beta S_i^\gamma  S_j^\delta\rangle &=
\lim_{n \to 0}  \left[ -(\beta J)^{-2} \delta_{\alpha\beta,\gamma\delta} +
\langle q_{\alpha\beta}\, q_{\gamma\delta}\rangle \right ] ,
\label{avs}
\end{align}
\end{subequations}
where the averages are to be evaluated with $\Delta_{\alpha\beta} = 0$.

Hence, from Eqs.~\eqref{chisg}, \eqref{chisg_terms} and \eqref{avs} we
have\footnote{Unfortunately, in Eq.~(4.47) of the review of
Binder and Young~\cite{binder:86}, 
which is the equation corresponding to our Eq.~\eqref{chisg_q_b},
the term $-(\beta J)^{-2}$ is missing.}
\begin{equation}
\chiSG = {1 \over T^2} \left[-(\beta J)^{-2} + \langle
q_{\alpha\beta}^2\rangle - 2 \langle q_{\alpha\beta}q_{\alpha\gamma}\rangle
+ \langle q_{\alpha\beta} q_{\gamma\delta}\rangle \right] \, .
\label{chisg_q_b}
\end{equation}
We write
\begin{equation}
q_{\alpha\beta} = q_c + \delta q_{\alpha\beta} 
\label{qc}
\end{equation}
where $q_c$ is the value of the $q_{\alpha\beta}$
at the (replica-symmetric) saddle point. Inserting
Eq.~\eqref{qc} into Eq.~\eqref{chisg_q_b} the factors of $q_c$ cancel and so we have
\begin{equation}
\chiSG = -{1 \over J^2}  + {1 \over T^2} \left[ \langle
\delta q_{\alpha\beta}^2\rangle - 2 \langle \delta q_{\alpha\beta}\delta q_{\alpha\gamma}\rangle
+ \langle \delta q_{\alpha\beta} \delta q_{\gamma\delta}\rangle \right] \, .
\label{chisg_q}
\end{equation}
The averages involve Gaussian integrals which come from the weight given by Eq.~\eqref{Zn2}
in which $f[\{q\}]$ is given by the quadratic expression in Eq.~\eqref{fq_quadratic}. 

We will need the result that if a set of variables $x_i$ have Gaussian distribution, i.e.
\begin{equation}
P(\{x\}) \propto \exp[-{\smfrac{1}{2} x_i A_{ij} x_j}] \, 
\end{equation}
then correlation functions of the $x_i$ are given by
\begin{equation}
\langle x_i x_j \rangle = \left(A^{-1}\right)_{ij} . 
\label{Ainv}
\end{equation}
The combination of averages in Eq.~\eqref{chisg_q} corresponds to the ``replicon''
eigenvector of the matrix $A$, see 
Eqs.~\eqref{abab}--\eqref{lar}. Hence, from Eq.~\eqref{Ainv}, these averages just give
the inverse of the replicon eigenvalue $\lambda_r$ in Eq.~\eqref{lambda_r} so
\begin{align}
\chiSG &=  -{1 \over J^2} + { 1 \over T^2} {1 \over \lambda_r}, \\
&=  -{1 \over J^2} + { 1 \over J^2}{1 \over 1 - (\beta J)^2(1 - 2 q + r) } ,
\\
&= \beta^2 { 1 - 2 q + r \over 1 - (\beta J)^2 (1 - 2 q + r) } \, .
\end{align}
If we define
\begin{equation}
\boxed{\chi^0_{SG} = \beta^2 (1 -2 q + r), }
\label{chisg0}
\end{equation}
then
\begin{equation}
\boxed{\chiSG = {\chi^0_{SG} \over 1 - J^2 \chi^0_{SG} }, }
\label{chisg_final}
\end{equation}
a result which is very reminiscent of the random phase approximation.

Equations \eqref{chisg0} and \eqref{chisg_final} are also valid in the
presence of a field, either uniform or random, provided the expressions for
$q$, $r$ and $\lambda_r$, in Eqs.~\eqref{q_sc},\eqref{abcd} and \eqref{lambda_r_b}
are modified appropriately. For
example, for a Gaussian random field with standard deviation $h$, the factor
of $J q^{1/2}$ is replaced by $(J^2 q + h^2)^{1/2}$, see
Refs.~\cite{sharma:10,singh:17b}, and for a unform field, $Jq^{1/2} z$ is
replaced by $Jq^{1/2} z + h$, see e.g.~Ref.~\cite{binder:86}.

In the paramagnetic phase, where $q=r=0$ and so $\chiSG^0 = \beta^2$, we see 
that $\chiSG$ has the simple form
\begin{equation}
\chiSG = {1 \over T^2 - J^2 }, \qquad (T > T_c = J) ,
\end{equation}
which shows that the transition occurs when $T = J$, as is well
known~\cite{sherrington:75}, see also Eq.~\eqref{Tc}.

\section{The Quantum SK model}
\label{sec:quant}

Now we make the model \textbf{quantum} by adding a transverse field. 
\begin{equation}
\mathcal{H} = -\sum_{\langle i, j \rangle} J_{ij} \sigma_i^z \sigma_j^z - h^T
\sum_i \sigma_i^x \, .
\end{equation}
This model has been studied in many works,
including
Refs.~\cite{yamamoto:87,ray:89,lai:90,goldschmidt:90,buttner:90,read:95,federov:86,ye:93,huse:93},
and these notes will use their methods.

The standard approach is to
use the imaginary time path integral formulation~\cite{sachdev:99},
where imaginary
time, $\tau$, is in the range $0 \le \tau \le \beta$ and there are periodic
boundary conditions in the $\tau$ direction. Imaginary time is divided
into $M$ time slices, each of width
\begin{equation}
\Delta \tau = {\beta \over M} \, .
\end{equation}
The partition function is then given by the action
\begin{equation}
Z = \mathrm{Tr} \exp \left[ \sum_{l=1}^M \left( \sum_{\langle i, j \rangle} J_{ij}
S_i(l) S_j(l) \Delta\tau + K^\tau \sum_i S_i(l)
S_i(l+1) \right) \right] \, ,
\end{equation}
where 
\begin{equation}
e^{-2 K^\tau} = \tanh(h^T \Delta\tau) \, ,
\label{Ktau}
\end{equation}
and the $S_i(l)$ are Ising variables at each site $i$ and time slice $l$. 
The $K^\tau$ term is a ferromagnetic coupling along the imaginary time
direction. Now we replicate, in order to average over disorder. Disorder averaging
does not alter the
$K^\tau$ term because it is not random, so averaging over the $J_{ij}$ term goes
through as for the classical case, but with the addition of the imaginary time indices. The
analog of Eq.~\eqref{Zn0} is
\begin{equation}
\begin{split}
\left[ Z^n \right]\av = \sum_{\{S_i^\alpha(l)\} = \pm 1}\, \exp
\bigg[
{(\Delta\tau J)^2 \over 2 N} \sum_{l,l'=1}^M \sum_{\langle i, j \rangle}
\sum_{\alpha,\beta = 1}^n 
S_i^\alpha(l) S_j^\alpha(l) S_i^\beta(l') S_j^\beta(l')
\ + \\
K^\tau \sum_i \sum_{l=1}^M \sum_{\alpha=1}^n S_i^\alpha(l)
S_i^\alpha(l+1) 
\bigg]  \, .
\end{split}
\label{Zn_quant}
\end{equation}
In the first term in the exponential we consider separately the $\alpha = \beta$
and $\alpha \ne \beta$ terms.
\begin{itemize}
\item $\alpha = \beta$ terms.\\
\begin{align}
& \sum_{l,l'=1}^M \sum_{\langle i, j \rangle}
\sum_{\alpha= 1}^n 
S_i^\alpha(l) S_j^\alpha(l) S_i^\alpha(l')
S_j^\alpha(l')
\\ &= 
{1 \over 2} \sum_{\alpha= 1}^n \sum_{l,l'=1}^M \left(\sum_i S_i^\alpha(l)
S_i^\alpha(l') \right)^2 \, ,
\end{align}
where we have neglected terms of order $1/N$. We will decouple the square
using a Hubbard-Stratonovich (HS) transformation as we did to go from
Eq.~\eqref{Zn1} to \eqref{Zn}.

\item $\alpha \ne  \beta$ terms.\\
\begin{equation}
\sum_{\alpha < \beta} \sum_{l,l'=1}^M \left(\sum_i S_i^\alpha(l)
S_i^\beta(l') \right)^2 \, ,
\end{equation}
We will do a HS transformation for this too.
\end{itemize}

The result of the HS transformations is
\begin{equation}
\begin{split}
\left[ Z^n \right]\av =
\prod_{\alpha,l,l'} & \left[
\sqrt{N \over 4 \pi} (\Delta\tau J)
\int_{-\infty}^\infty
d r_\alpha(l, l')\right]
\prod_{\alpha<\beta,l,l'}\left[
\sqrt{N \over 2 \pi} (\Delta\tau J)
\int_{-\infty}^\infty
d q_{\alpha\beta}(l, l')\right] \\
&\exp \biggl[ -{N \over 4} (\Delta\tau J)^2 \sum_{l,l',\alpha} r_\alpha^2(l,l')
-{N \over 2} (\Delta\tau J)^2 \sum_{l,l',\alpha< \beta}
q_{\alpha\beta}^2(l,l') 
\biggr] 
\left(\mathrm{Tr}\, e^{-H}\right)^N \, ,
\end{split}
\end{equation}
where
\begin{equation}
\begin{split}
H = -(\Delta\tau J)^2  & \left[ {1 \over 2}
\sum_{\alpha}\sum_{l, l'} r_\alpha(l,l') S^\alpha(l) S^\alpha(l') +
\sum_{\alpha< \beta}\sum_{l, l'} q_{\alpha\beta}(l,l') S^\alpha(l) S^\beta(l')
\right] \\
- &
K^\tau \sum_\alpha \sum_l S^\alpha(l) S^\alpha(l + 1) \, .
\end{split}
\end{equation}

We have time translational invariance so $q_{\alpha\beta}(l, l')$ is only a
function of $\Delta l \equiv l - l'$, and similarly for $r_{\alpha}(l, l')$.
Hence
\begin{equation}
\label{Znh}
\begin{split}
\left[ Z^n \right]\av =
\prod_{\alpha,\Delta l} & \left[
\sqrt{N \over 2 \pi} (\Delta\tau J)
\int_{-\infty}^\infty
d r_\alpha(\Delta l)\right]
\prod_{\alpha<\beta,\Delta l}\left[
\sqrt{N \over 2 \pi} (\Delta\tau J)
\int_{-\infty}^\infty
d q_{\alpha\beta}(\Delta l)\right] \\
&\exp \biggl[ -{N \over 4} (\Delta\tau J)^2
M \sum_{\Delta l,\alpha} r_\alpha^2(\Delta l)
-{N \over 2} (\Delta\tau J)^2 M \sum_{\Delta l,\alpha< \beta}
q_{\alpha\beta}^2(\Delta l) 
\biggr] 
\left(\mathrm{Tr}\, e^{-H}\right)^N \, ,
\end{split}
\end{equation}
where
\begin{equation}
\begin{split}
H = -(\Delta\tau J)^2  & \left[ {1 \over 2}
\sum_{\alpha}\sum_{\Delta l} r_\alpha(\Delta l)
\sum_l S^\alpha(l) S^\alpha(l+\Delta l) +
\sum_{\alpha< \beta}\sum_{\Delta l} q_{\alpha\beta}(\Delta l)
\sum_l S^\alpha(l) S^\beta(l+\Delta l)
\right] \\
- &
K^\tau \sum_\alpha \sum_l S^\alpha(l) S^\alpha(l + 1) \, .
\end{split}
\label{L_qu}
\end{equation}
We minimize w.r.t.~$q_{\alpha\beta}(\Delta l)$ and $r_{\alpha}(\Delta l)$. This
gives 
\begin{subequations}
\begin{align}
r_{\alpha}(\Delta l) &= \langle S^\alpha(0)
S^\alpha(\Delta l) \rangle \, ,
\\
q_{\alpha\beta}(\Delta l) &= \langle
S^\alpha(0) S^\beta(\Delta l)
\rangle ,
\end{align}
\end{subequations}
where the average is with respect to $e^{-H}$ and we have used time translational invariance. 

We write Eq.~\eqref{Znh} as
\begin{equation}
\label{Znf}
\left[ Z^n \right]\av =
\prod_{\alpha,\Delta l}  \left[
\sqrt{N \over 2 \pi} (\Delta\tau J)
\int_{-\infty}^\infty
d r_\alpha(\Delta l)\right]
\prod_{\alpha<\beta,\Delta l}\left[
\sqrt{N \over 2 \pi} (\Delta\tau J)
\int_{-\infty}^\infty
d q_{\alpha\beta}(\Delta l)\right] 
\exp[-Nn \beta f]
\end{equation}
where
\begin{equation}
-\beta f = \lim_{n\to0} \left[
-{1 \over 4 n} \beta\Delta \tau J^2 \sum_{\alpha,\Delta l} r_\alpha(\Delta l)^2
-{1 \over 2 n} \beta\Delta \tau J^2 \sum_{\alpha<\beta} \sum_{\Delta l}
q_{\alpha\beta}^2(\Delta l) 
+ {1 \over n} \, \ln 
\mathrm{Tr}\, e^{-H}\, \right].
\label{f_qu}
\end{equation}
We evaluate the integrals in Eq.~\eqref{Znf} by steepest descent and 
look for the replica symmetric solution:
\begin{subequations}
\begin{align}
r_\alpha(\Delta l) &= r(\Delta l) \, , \\
q_{\alpha\beta}(\Delta l) &= q(\Delta l) \, .
\end{align}
\end{subequations}
This yields
\begin{equation}
-\beta f = \lim_{n\to0} \left[
-{1 \over 4} \beta\Delta \tau J^2 \sum_{\Delta l} r(\Delta l)^2
+{1 \over 4} \beta\Delta \tau J^2 \sum_{\Delta l}
q^2(\Delta l) 
+ {1 \over n} \, \ln 
\mathrm{Tr}\, e^{-H}\, \right],
\label{f1}
\end{equation}
where
\begin{equation}
\begin{split}
H = -(\Delta\tau J)^2  & \left[ {1 \over 2}
\sum_{\Delta l} r(\Delta l)
\sum_\alpha \sum_l S^\alpha(l) S^\alpha(l+\Delta l) +
\sum_{\Delta l}
q(\Delta l)
\sum_{\alpha< \beta}
\sum_l S^\alpha(l) S^\beta(l+\Delta l)
\right] \\
- &
K^\tau \sum_\alpha \sum_l S^\alpha(l) S^\alpha(l + 1) \, .
\end{split}
\label{L_a}
\end{equation}

Now we need to think about the physics. The parameters $q(\Delta l)$ are order
parameters, corresponding to a product of a single spin in two replicas. We
expect these to be independent of time. Hence we will assume that
$q(\Delta l) = q$.  Consequently
\begin{equation}
-\beta f = \lim_{n\to0} \left[
-{1 \over 4} \beta\Delta \tau J^2 \sum_{\Delta l} r(\Delta l)^2
+{1 \over 4} (\beta J)^2 
q^2
+ {1 \over n} \, \ln 
\mathrm{Tr}\, e^{-H}\, \right] .
\end{equation}
We write the second term in the expression for $H$ in Eq.~\eqref{L_a} as
follows
\begin{equation}
{1\over 2} (\Delta\tau J)^2 q \sum_{l_1,l_2} \sum_{(\alpha,\beta)}
S^\alpha(l_1) S^\beta(l_2) = 
{1\over 2} (\Delta\tau J)^2 q \left[ \left( \sum_{l} \sum_\alpha S_\alpha(l)
\right)^2 
-\sum_\alpha \sum_{l,\Delta l} S^\alpha(l) S^\alpha(l+\Delta l)
\right] .
\label{xx}
\end{equation}
The second term on the RHS of Eq.~\eqref{xx} can be combined with the $r$ term
in Eq.~\eqref{L_a}. The $K^\tau$ term in Eq.~\eqref{L_a} can also be combined 
so we define
\begin{align}
\overline{r}(\Delta l) &= r(\Delta l) - q\, ,
\qquad\qquad\qquad\qquad\ (\Delta l \ne 1)\, ,
\nonumber \\
\overline{r}(\Delta l) &= r(\Delta l) - q  +K^\tau /(\Delta\tau J)^2 \, .
\qquad (\Delta l = 1) \, ,
\label{or_r}
\end{align}
The square in the first term on the RHS of Eq.~\eqref{xx}
is decoupled by a Hubbard-Stratonovich transformation, as in the
classical case, see Eqs.~\eqref{zSa_all}, i.e.
\begin{align}
& \mathrm{Tr} \exp\Biggl[
{(\Delta\tau J)^2\over 2} \left( 
\sum_{\Delta l} \overline{r}(\Delta l)
\sum_\alpha \sum_l S^\alpha(l) S^\alpha(l+\Delta l) +
q \left( \sum_l \sum_\alpha S^\alpha(l) \right)^2 \right) 
\Biggr] \nonumber \\
& =  {1\over \sqrt{2 \pi}} \int_{-\infty}^\infty dz e^{-z^2/2}
\mathrm{Tr} \exp\Biggl[ \sum_\alpha\left(
{(\Delta\tau J)^2\over 2} 
\sum_{\Delta l} \overline{r}(\Delta l)
\sum_l S^\alpha(l) S^\alpha(l+\Delta l) +
(\Delta \tau J) q^{1/2} z \sum_l S^\alpha(l)  \right)
\Biggr]  \nonumber \\
&={1\over \sqrt{2 \pi}} \int_{-\infty}^\infty dz e^{-z^2/2}
\left(
\mathrm{Tr}\, e^{-\overline{H}(z)}
\right)^n
\label{Z1n}
\end{align}
in which $\overline{H}(z)$ is given by
\begin{equation}
\boxed{
\overline{H}(z) = -(\Delta\tau J)^2 
\sum_{\langle l_1, l_2\rangle} \left\{\,r(|l_1 - l_2|) - q\, \right\}
S(l_1) S(l_2) - K^\tau \sum_l S(l) S(l+1) -
(\Delta \tau J) q^{1/2} z \sum_l S(l) \, ,}
\label{olineL}
\end{equation}
where we used Eq.~\eqref{or_r}.
We see that 
$\overline{H}(z)$ 
is the Hamiltonian of a one-dimensional chain with long-range interactions,
in which there is a
(uniform) field proportional to $q^{1/2} z$.

Expanding Eq.~\eqref{Z1n} in
powers of $n$ and substituting into Eq.~\eqref{f1} gives
\begin{equation}
\boxed{
-\beta f = \lim_{n\to0} \left[
-{1 \over 4} \beta\Delta \tau J^2 \sum_{\Delta l = 1}^M r(\Delta l)^2
+{1 \over 4} (\beta J)^2 q^2
+ {1 \over \sqrt{2 \pi}} \int_{-\infty}^\infty dz \, e^{-z^2/2} \,
\ln \mathrm{Tr}\, e^{-\overline{H}(z)} \right].}
\end{equation}

Now we determine the self-consistent equations for $r(\Delta \tau)$ and $q$. 
\begin{itemize}
\item $r(\Delta l)$. \\
\begin{equation}
{1\over 2} \, \beta\Delta\tau J^2 r(\Delta l) = {1 \over \sqrt{2 \pi}} 
\int_{-\infty}^\infty dz\,e^{-z^2/2}\, \left\{
{(\Delta\tau J)^2 \over 2}{\mathrm{Tr}\, \sum_l S(l) S(l+\Delta l) 
e^{-\overline{H}(z)} \over \mathrm{Tr}\, e^{-\overline{H}(z)} } \, ,
\right\}
\end{equation}
so
\begin{equation}
\boxed{
r(\Delta l) = {1 \over \sqrt{2 \pi}} 
\int_{-\infty}^\infty dz\,e^{-z^2/2}\, 
\langle \, S(0) \, S(\Delta l) \,\rangle_{\overline{H}} \, ,}
\label{r_sc}
\end{equation}
where we have used time translational invariance and
\begin{center}
\fbox{$\langle \cdots
\rangle_{\overline{H}}$ indicates an average over the spins with weight
$e^{-\overline{H}}$.} 
\end{center}
\item $q$.
\begin{multline}
{1\over 2}(\beta J)^2 q = -{1 \over \sqrt{2 \pi}} 
\int_{-\infty}^\infty dz\,e^{-z^2/2}\, \times \\
\left\{
{
{(\Delta\tau J) \over 2 q^{1/2}}\, z\, \mathrm{Tr}\, \left(
\sum_l S(l) e^{-\overline{H}(z)}\right)
-{(\Delta\tau J)^2 \over 2} \mathrm{Tr}\, \sum_{l,\Delta l}
S(l) S(l+\Delta l)
e^{-\overline{L}(z)}
\over \mathrm{Tr}\, e^{-\overline{H}(z)}
}
\right\}
\end{multline}
\end{itemize}
As in the classical case, we integrate by parts with respect to $z$ in the first
term in curly brackets. This gives
\begin{multline}
-(\beta J)^2 q = {1 \over \sqrt{2 \pi}} 
\int_{-\infty}^\infty dz\,e^{-z^2/2}\, \times \\
\left\{
{(\Delta\tau J) \over q^{1/2}}\, (\Delta \tau J) q^{1/2}\,
\sum_{l_1, l_2}
\biggl( 
\langle S(l_1) S(l_2) \rangle_{\overline{H}}
- \langle S(l_1) \rangle_{\overline{H}}\, \langle
S(l_2) \rangle_{\overline{H}}
\biggr)
-(\Delta\tau J)^2  \sum_{l_1,l_2} 
\langle S(l_1) S(l_2) \rangle_{\overline{H}}
\right\} \, ,
\end{multline}
which simplifies to
\begin{equation}
\boxed{
q = {1 \over \sqrt{2 \pi}} 
\int_{-\infty}^\infty dz\,e^{-z^2/2}\, \langle \, S(0)
\, \rangle^2_{\overline{H}}\, 
 \, ,
}
\label{q_sc_qu}
\end{equation}
where the average of a single spin $\langle S(0) \rangle_{\overline{H}}$
could be evaluated at any time slice because of time translational invariance.

We now check that we recover the standard results for the SK model when we
take the classical limit $h^T \to 0$. From Eq.~\eqref{Ktau}
this corresponds to $K^\tau \to \infty$.
Hence all spins along the time direction are fully correlated. It follows
that
\begin{equation}
\langle S(0) S(\Delta l) \rangle = 1, \qquad \text{for
all\ }\Delta l, 
\end{equation}
and so, from Eq.~\eqref{r_sc},
\begin{equation}
r(\Delta l) = 1,  \qquad \text{for
all\ }\Delta l.
\end{equation}
In $\overline{H}(z)$ in Eq.~\eqref{olineL} the first two terms are constants
which
cancel when computing averages. In the third term all the $S(l)$ are
equal and so we can write $\sum_l S(l) = M S$ where $S = \pm 1$, so
\begin{equation}
(\Delta \tau J)q^{1/2} z \sum_l S(l) = (\beta J) q^{1/2} z S, 
\end{equation}
exactly in the classical case, Eq.~\eqref{expH}. Hence
\begin{equation}
\langle S(l)\rangle_{\overline{H}} = 
\langle S\rangle_{\overline{H}} = \tanh (\beta J q^{1/2} z) \, ,
\end{equation}
and the self-consistent equation of $q$, Eq.~\eqref{q_sc_qu}, reduces to the
result for the SK model, Eq.~\eqref{q_sc}.

We now consider the stability of the replica symmetric solution in the quantum
case. The free energy is given by Eq.~\eqref{f_qu} (but with $q_{\alpha\beta}$ now
independent of $\Delta l$) and $H$ given by Eq.~\eqref{L_qu}. The stability has
to be determined with respect to the $n(n-1)/2$ static order parameters $q_{\alpha\beta}$
and the $n M$ correlation functions $r_\alpha(\Delta l)$.
The matrix of
second derivatives therefore has the form
\begin{equation}
T = \begin{pmatrix}
A & B \\
B^T & C \\
\end{pmatrix} \, ,
\label{T}
\end{equation}
where $A$ is a matrix of dimension $n(n-1)/2 \times n(n-1)/2$ which describes
fluctuations in the $q_{\alpha\beta}$ sector, $C$ is of size $n M \times n M$
and describes fluctuations in the $r_\alpha(\Delta l)$ sector, and $B$, which is of dimension
$n(n-1)/2\times n M$, describes the mixed second derivatives. 

For the classical case, i.e.\ the SK model, de Almeida and
Thouless (AT)~\cite{almeida:78} showed that
the instability comes in the ``replicon'' eigenvector of $A$, see
Eq.~\eqref{lambda_r}. AT also showed that, in cases where one also has to
consider terms diagonal in replica indices (such as the $r$ terms here) the
replicon eigenvector has zero values for these components. Here we assume that
the replicon mode will still be the important one, and so we will neglect the
sector involving the $r$ terms~\cite{buttner:90}.
Thus we just need to consider the $n(n-1)/2
\times n(n-1)/2$ matrix $A$, and so write the expansion of the free energy as in
Eq.~\eqref{fq_quadratic}.

Taking the derivatives in Eq.~\eqref{f_qu}, we find that
$A^{\alpha\beta,\gamma\delta}$, the matrix of
coefficients in the expansion in Eq.~\eqref{fq_quadratic}, is given by
\begin{align}
A^{\alpha\beta,\gamma\delta} & \equiv {\partial^2  (\beta f) \over \partial
q_{\alpha\beta} \partial q_{\gamma\delta} } \ \ (\times n) \, , \nonumber  \\
&= J^2 \beta \Delta \tau M \delta_{\alpha\beta,\gamma\delta} - (\beta \Delta
\tau)^4 \sum_{l_1,l_2,l_3,l_4} \left[
\langle S^\alpha(l_1) S^\beta(l_2)
S^\gamma(l_3) S^\delta(l_4)\rangle - \langle S^\alpha(l_1)
S^\beta(l_2)\rangle \langle S^\gamma(l_3) S^\delta(l_4)\rangle
\right] \,  \nonumber \\
&= (\beta J)^2 \delta_{\alpha\beta,\gamma\delta} - (\beta J)^4 \left[
\left({1 \over M^4}\,
\sum_{l_1,l_2,l_3,l_4} 
\langle S^\alpha(l_1) S^\beta(l_2)
S^\gamma(l_3) S^\delta(l_4)\rangle \right) -
q^2 \right] \, ,
\label{Aab_qu}
\end{align}
where we used that $\langle S^\alpha(l_1) S^\beta(l_2)\rangle$
for $\alpha \ne \beta$
is independent of the values of $\tau$ and is just order the order parameter
$q_{\alpha\beta} $( $= q$ here since we expand about the replica
symmetric solution). Eq.~\eqref{Aab_qu} is the generalization to the quantum case of
Eq.~\eqref{Aab}.

As for the classical case, we have to consider three cases depending on which
replica indices are equal, see Eq.~\eqref{abab}, \eqref{abac} and \eqref{abcd}.
The averages are to be evaluated in the replica symmetric solution, so the
spin averages in each replica are to be evaluated with weight
$e^{-\overline{H}(z)}$, where $\overline{H}(z)$ is given by Eq.~\eqref{olineL}, and
finally the combined average over the different replicas
is to be averaged over the Gaussian random field $z$ which has zero mean and standard
deviation unity, see e.g.~Eq.~\eqref{q_sc_qu} which is for an average over
two different replicas.

\begin{itemize}
\item $(\alpha\beta) (\alpha\beta)$. \\
\begin{equation}
\left({1 \over M^4}\,
\sum_{l_1,l_2,l_3,l_4} 
\langle S^\alpha(l_1) S^\beta(l_2)
S^\alpha(l_3) S^\beta(l_4)\rangle \right) =
{1 \over \sqrt{2 \pi}}\int_{-\infty}^\infty  dz\,e^{-z^2/2} \,
\left(
{1 \over M^2}
\sum_{l_1, l_2}
\langle\, S(l_1) S(l_2)\, \rangle_{\overline{H}} \right)^2 \, .
\end{equation}

\item $(\alpha\beta) (\alpha\gamma)$ with $\beta \ne \gamma$. \\
\begin{equation}
\begin{split}
& \left({1 \over M^4}\, 
\sum_{l_1,l_2,l_3,l_4} 
\langle S^\alpha(l_1) S^\beta(l_2)
S^\alpha(l_3) S^\gamma(l_4)\rangle \right) = \\
& \qquad\qquad {1 \over \sqrt{2 \pi}}\int_{-\infty}^\infty  dz\,e^{-z^2/2} \,
\left(
{1 \over M^2}
\sum_{l_1, l_2}
\langle\, S(l_1) S(l_2)\, \rangle_{\overline{H}} \right)
\left(
{1 \over M}
\sum_{l}
\langle\, S(l)\, \rangle_{\overline{H}} \right)^2 \, .
\end{split}
\end{equation}

\item $(\alpha\beta) (\gamma\delta)$ with all indices different. \\
\begin{equation}
\left({1 \over M^4}\,
\sum_{l_1,l_2,l_3,l_4} 
\langle S^\alpha(l_1) S^\beta(l_2)
S^\gamma(l_3) S^\delta(l_4)\rangle \right) =
{1 \over \sqrt{2 \pi}}\int_{-\infty}^\infty dz\, e^{-z^2/2} \,
\left( {1 \over M} \sum_l \langle \, S(l) \, \rangle_{\overline{H}} \right)^4 \, .
\label{abgd_qn}
\end{equation}

\end{itemize}
As in the classical case, defining
$(\beta J)^{-2} A^{\alpha\beta,\alpha\beta} = P$, 
$(\beta J)^{-2} A^{\alpha\beta,\alpha\gamma} = Q$, and
$(\beta J)^{-2} A^{\alpha\beta,\gamma\delta} = R$, the replicon eigenvalue is given by
\begin{align}
(\beta J)^{-2} \lambda_r &= P - 2 Q + R \nonumber \\
& \boxed{\,= 1 - J^2 \, {1 \over \sqrt{2 \pi}}\, \int_{-\infty}^\infty
dz \, e^{-z^2/2} \,  \left[ \sum_l \Bigl( \Delta \tau \{\langle \,
S(0) S(l)\, \rangle_{\overline{H}} - \langle \,S(0)
\,\rangle_{\overline{H}} \, \langle
\, S(l) \,\rangle_{\overline{H}} \} \Bigr) \right]^2 \, .}
\label{lambda_r_qu}
\end{align}
The correspondence with the classical (SK model) result in
Eq.~\eqref{lambda_r_b} is clear.

To summarize,
averages denoted by $\langle \cdots \rangle_{\overline{H}}$ are
with respect to weight $e^{-\overline{H}(z)}$ where $\overline{H}(z)$ is given
by Eq.~\eqref{olineL}.
The $M-1$ values of $r(\Delta l)$, as well as $q$, are to be
determined self-consistently from Eqs.~\eqref{r_sc} and \eqref{q_sc_qu}.

In
the continuum limit the expression corresponding to Eq.~\eqref{lambda_r_qu} is
clearly
\begin{equation}
(\beta J)^{-2} \lambda_r = 
1 - J^2 \, {1 \over \sqrt{2 \pi}}\, \int_{-\infty}^\infty
dz \, e^{-z^2/2} \,\left[
\int_0^\beta d \tau \Bigl( \langle \,
S(0) S(\tau)\, \rangle_{\overline{H}} - \langle \,S(0)
\,\rangle_{\overline{H}} \, \langle
\, S(\tau) \,\rangle_{\overline{H}} \Bigr) \right]^2 \, .
\end{equation}

In the paramagnetic phase, single spin expecation values vanish, and so does
$q$, and the integral over $z$ gives unity, so
\begin{equation}
\boxed{
\lambda_r = (\beta J)^2 \left\{1 - 
\left[
\sum_{l} \Bigl( \Delta \tau \langle \,
S(0) S(\tau)\, \rangle_{\overline{H}} 
\Bigr) \right]^2 \right\} \qquad (\text{where\ } q = 0) .}
\end{equation}
The phase boundary is where $\lambda_r = 0$, i.e.
\begin{equation}
\boxed{
1 = J \Delta\tau \sum_l \langle S(0) S(l)\, \rangle_{\overline{H}}
\qquad (\mathrm{on\ phase\ boundary})\  ,}
\label{critline}
\end{equation}
where now, since $q=0$,
\begin{equation}
\boxed{
\overline{H} = 
- (\Delta\tau J)^2
\sum_{\langle l_1, l_2\rangle} r(|l_1 - l_2|)\,
S(l_1) S(l_2) - K^\tau \sum_l S(l)S(l+1) , \qquad (\text{if}\ q = 0),}
\end{equation}
see Eq.~\eqref{olineL}.

As for the classical case we want to calculate the spin glass susceptibility,
since the divergence of this quantity is governed by the vanishing of the
replicon eigenvalue. In the quantum case, $\chiSG$ is given by
\begin{equation}
\chiSG = {1 \over N} \sum_{i, j=1}^N \left[ \left(
\int_0^\beta [\langle \sigma^z_i(\tau) \sigma^z_j(0) \rangle -
\langle \sigma^z_i\rangle \langle \sigma^z_j\rangle]
\, d\tau
\right)^2 \right]\av .
\label{chisg_qu}
\end{equation}
In the path integral formulation, this becomes
\begin{equation}
\chiSG = {1 \over N} \sum_{i, j=1}^N \left[ \left(
\sum_{l=1}^M  [\langle S_i(l_0+l) S_j(l_0) \rangle -
\langle S_i(l_0)\rangle \langle S_j(l_0)\rangle]
\, \Delta \tau
\right)^2 \right]\av ,
\label{chisg_qu_dtau}
\end{equation}
where we have discretized imaginary time as before.

We consider each term separately, and average over all possible time slices,
which means sum over four time labels and have to divide by $M^2$. We follow
similar steps to those in the classical case in Sec.~\ref{sec:class}.
\begin{subequations}
\label{chisg_terms_qu}
\begin{multline}
{1 \over M^2} \sum_{l_1, l_2, l_3, l_4=1}^M
\langle S_i(l_1)  S_j(l_2) \rangle\, \langle S_i(l_3)  S_j(l_4) \rangle
= \\
{1 \over M^2}\lim_{n \to 0} {1 \over n(n-1)} \sum_{(\alpha, \beta)}
\sum_{l_1, l_2, l_3, l_4=1}^M 
\langle S_i^\alpha(l_1)  S_j^\alpha(l_2) S_i^\beta(l_3)  S_j^\beta(l_4) \rangle ,
\end{multline}
\begin{multline}
{1 \over M^2} \sum_{l_1, l_2, l_3, l_4=1}^M
\langle S_i(l_1)  S_j(l_2) \rangle\, \langle S_i(l_3)\rangle \,
\langle S_j(l_4) \rangle
= \\
{1 \over M^2}\lim_{n \to 0} {1 \over n(n-1)(n-2)} \sum_{(\alpha,\beta,\gamma)}
\sum_{l_1, l_2, l_3, l_4=1}^M 
\langle S_i^\alpha(l_1)  S_j^\alpha(l_2) S_i^\beta(l_3)  S_j^\gamma(l_4) \rangle ,
\end{multline}
\begin{multline}
{1 \over M^2} \sum_{l_1, l_2, l_3, l_4=1}^M
\langle S_i(l_1) \rangle\,\langle S_j(l_2) \rangle\, \langle S_i(l_3)\rangle \,
\langle S_j(l_4) \rangle
= \\
{1 \over M^2}\lim_{n \to 0} {1 \over n(n-1)(n-2)(n-3)}
\sum_{(\alpha,\beta,\gamma,\delta)}
\sum_{l_1, l_2, l_3, l_4=1}^M 
\langle S_i^\alpha(l_1)  S_j^\beta(l_2) S_i^\gamma(l_3)  S_j^\delta(l_4) \rangle ,
\end{multline}
\end{subequations}
where the sums $(\alpha,\beta,\gamma)$ etc.~are over all distinct pairs of replicas.

How do we do the averages in Eqs.~\eqref{chisg_terms_qu}? We refer to
Eq.~\eqref{Zn_quant} which we reproduce again here, in a slightly modified
form.
\begin{multline}
\left[ Z^n \right]\av = \sum_{\{S_i^\alpha(l)\} = \pm 1}\, \exp
\bigg[
{(\Delta\tau J)^2 \over 2 N} \sum_{l,l'=1}^M \sum_{\langle i, j \rangle}
\bigg(\sum_{(\alpha,\beta)}
S_i^\alpha(l) S_j^\alpha(l) S_i^\beta(l') S_j^\beta(l') + \\
+ \sum_\alpha S_i^\alpha(l) S_j^\alpha(l) S_i^\alpha(l') S_j^\alpha(l')
\bigg)
+ 
K^\tau \sum_i \sum_{l=1}^M \sum_{\alpha=1}^n S_i^\alpha(l)
S_i^\alpha(l+1) 
\bigg]  \, .
\label{Zn_quant_2}
\end{multline}
We now add fictitious fields $\Delta_{\alpha\beta}$ in 
in Eq.~\eqref{Zn_quant_2}, i.e.~we add
\begin{equation}
{1 \over M^2} \sum_{\alpha < \beta} \Delta_{\alpha\beta} \sum_{i=1}^N
\sum_{l_1, l_2 = 1}^M S_i^\alpha(l_1)   S_i^\beta(l_2) 
\end{equation}
in the exponent.
Remembering that, for $n\to0$, there is no normalizing denominator, we have
\begin{subequations}
\label{derivs_qu}
\begin{align}
{1 \over M^2} \sum_i \sum_{l_1, l_2=1}^M 
\langle S_i^\alpha(l_1) S_i^\beta(l_2) \rangle
&=
\lim_{n\to 0} {\partial \over \partial
\Delta_{\alpha\beta}}\left[ Z^n \right]\av, \\
{1 \over M^4} \sum_{i,j} \sum_{l_1, l_2, l_3, l_4=1}^M 
\langle S_i^\alpha(l_1) S_i^\beta(l_2) S_j^\gamma(l_3) S_j^\delta(l_4) \rangle &=
\lim_{n\to 0}
{\partial^2 \over \partial \Delta_{\alpha\beta} \Delta_{\gamma\delta}}
\left[ Z^n \right]\av, 
\end{align}
\end{subequations}
in which the replicas $\alpha,\beta,\gamma,\delta$ can take any values subject
to the restrictions $\alpha < \beta, \gamma < \delta$.
Proceeding as earlier in this section, we get to Eq.~\eqref{Znf} but with the
$\Delta_{\alpha\beta}$ added, and also assume that $q_{\alpha\beta}(\Delta l)$
is independent of $\Delta l$ for $\alpha \ne \beta$, a result which we assumed
above but at a later stage. This gives
\begin{equation}
\begin{split}
\left[ Z^n \right]\av =
\prod_{\alpha,\Delta l} & \left[
\sqrt{N \over 2 \pi} (\Delta\tau J)
\int_{-\infty}^\infty
d r_\alpha(\Delta l)\right]
\prod_{\alpha<\beta}\left[
\sqrt{N \over 2 \pi} (\Delta\tau J)
\int_{-\infty}^\infty
d q_{\alpha\beta}\right] \\
&\exp \biggl[ -{N \over 4} (\Delta\tau J)^2
M \sum_{\Delta l,\alpha} r_\alpha^2(\Delta l)
-{N \over 2} (\Delta\tau J)^2 M^2 \sum_{\alpha< \beta}
q_{\alpha\beta}^2 
\biggr] 
\left(\mathrm{Tr}\, e^{-H}\right)^N \, ,
\end{split}
\label{ZnH_2}
\end{equation}
where
\begin{equation}
\begin{split}
H = -(\Delta\tau J)^2  & \Biggl[ {1 \over 2}
\sum_{\alpha}\sum_{\Delta l} r_\alpha(\Delta l)
\sum_l S^\alpha(l) S^\alpha(l+\Delta l) + \\
& \sum_{\alpha< \beta}
\sum_{l_1, l_2=1}^M
\biggl(q_{\alpha\beta} - (\beta J)^{-2} \Delta_{\alpha\beta}\biggr)
S^\alpha(l_1) S^\beta(l_2)
\Biggr] 
- 
K^\tau \sum_\alpha \sum_l S^\alpha(l) S^\alpha(l + 1) \, .
\end{split}
\label{L_qu2}
\end{equation}
As in the classical case we define shifted variables by
\begin{equation}
u_{\alpha\beta} = q_{\alpha\beta} - (\beta J)^{-2} \Delta_{\alpha\beta} \, ,
\end{equation}
in terms of which
\begin{equation}
\begin{split}
\left[ Z^n \right]\av = &
\prod_{\alpha,\Delta l} \left[
\sqrt{N \over 2 \pi} (\Delta\tau J)
\int_{-\infty}^\infty
d r_\alpha(\Delta l)\right]
\prod_{\alpha<\beta}\left[
\sqrt{N \over 2 \pi} (\Delta\tau J)
\int_{-\infty}^\infty
d q_{\alpha\beta}\right] \\
&\exp \biggl[ -{N \over 4} (\Delta\tau J)^2
M \sum_{\Delta l,\alpha} r_\alpha^2(\Delta l)
-{N \over 2} \sum_{\alpha< \beta}
\biggl\{
(\beta J)^2 u_{\alpha\beta}^2  + 2 \Delta_{\alpha\beta} u_{\alpha\beta} +
(\beta J)^{-2} \Delta_{\alpha\beta}^2
\biggr\}
\biggr]\times  \\
&\left(\mathrm{Tr}\, e^{-H}\right)^N \, ,
\end{split}
\label{ZnH_3}
\end{equation}
where
\begin{equation}
\begin{split}
H = -(\Delta\tau J)^2  & \Biggl[ {1 \over 2}
\sum_{\alpha}\sum_{\Delta l} r_\alpha(\Delta l)
\sum_l S^\alpha(l) S^\alpha(l+\Delta l) + \\
& \sum_{\alpha< \beta}
\sum_{l_1, l_2=1}^M u_{\alpha\beta} S^\alpha(l_1) S^\beta(l_2)
\Biggr] 
- 
K^\tau \sum_\alpha \sum_l S^\alpha(l) S^\alpha(l + 1) \, .
\end{split}
\label{L_qu3}
\end{equation}
Doing the derivatives in Eqs.~\eqref{derivs_qu}
gives 
\begin{subequations}
\label{derivs_qu_2}
\begin{align}
{1\over N}\, {1 \over M^2} \sum_i \sum_{l_1, l_2=1}^M 
\langle S_i^\alpha(l_1) S_i^\beta(l_2) \rangle
&= \lim_{n\to 0} \langle q_{\alpha\beta} \rangle,
\\
{1 \over N}\,{1 \over M^4} \sum_{i,j} \sum_{l_1, l_2, l_3, l_4=1}^M 
\langle S_i^\alpha(l_1) S_i^\beta(l_2) S_j^\gamma(l_3) S_j^\delta(l_4) \rangle
&=
\lim_{n\to 0} \left[-(\beta J)^{-2}\delta_{\alpha\beta, \gamma\delta} + 
\langle q_{\alpha\beta}\, q_{\gamma\delta}\rangle \right] ,
\label{avs_qu}
\end{align}
\end{subequations}
where the averages are to be evaluated with $\Delta_{\alpha\beta} = 0$.
Notice the strong similarity between Eq.~\eqref{avs_qu} and the corresponding one
for the classical case, Eqs.~\eqref{avs}.

Referring to Eqs.~\eqref{chisg_qu_dtau},
\eqref{chisg_terms_qu} and \eqref{avs_qu}, we see that $\chiSG$ is given by
\begin{equation}
\chiSG = \beta^2 \left[ (-\beta J)^{-2}  + 
\langle q_{\alpha\beta}^2\rangle - 2 \langle
q_{\alpha\beta}\,q_{\alpha\gamma}\rangle + \langle
q_{\alpha\beta}\,q_{\gamma\delta}\rangle \right] \, ,
\label{chisgdq}
\end{equation}
where we used that $M \Delta \tau = \beta$. Note that Eq.~\eqref{chisgdq} is
identical to the corresponding classical result, Eq.~\eqref{chisg_q_b}.
Separating out the replica symmetric saddle point value $q_c$ as in
Eq.~\eqref{chisg_q_b} we get
\begin{equation}
\chiSG =
-{1\over J^2}  + {1 \over T^2}
\left[
\langle \delta q_{\alpha\beta}^2\rangle - 2 \langle
\delta q_{\alpha\beta}\,\delta q_{\alpha\gamma}\rangle + \langle
\delta q_{\alpha\beta}\,\delta q_{\gamma\delta}\rangle \right] \, .
\label{chisgddq}
\end{equation}
The averages in Eq.~\eqref{chisgddq} are over Gaussian integrals given by the
weight in Eq.~\eqref{Znf} in which $f$ is given by the quadratic expression in
Eq.~\eqref{fq_quadratic}.  The combination of averages in Eq.~\eqref{chisgdq}
just corresponds to the ``replicon'' eigenvector of the matrix $A$, see
Eq.~\eqref{lambda_r_qu} and 
Eqs.~\eqref{Aab_qu}--\eqref{abgd_qn}. Hence, according to Eq.~\eqref{Ainv},
these averages just give the inverse of the
replicon eigenvalue in Eq.~\eqref{lambda_r_qu}. Consequently the spin glass
susceptibility is given, in the quantum case, by
\begin{align}
\chiSG & =  -{1 \over J^2} + {1 \over T^2} {1 \over \lambda_r}  , \\
&= -{1 \over J^2} + {1 \over J^2} {1 \over 1 - J^2 \chiSG^0} , \\
& \boxed{= {\chiSG^0 \over 1 - J^2 \chiSG^0} ,}
\label{chisg_chisg0}
\end{align}
where
\begin{equation}
\boxed{
\chiSG^0 = 
{1 \over \sqrt{2 \pi}}\, \int_{-\infty}^\infty
dz \, e^{-z^2/2} \,  \left[ \sum_l \Delta \tau \Bigl( \langle \,
S(0) S(l)\, \rangle_{\overline{H}} - \langle \,S(0)
\,\rangle_{\overline{H}} \, \langle
\, S(l) \,\rangle_{\overline{H}} \Bigr) \right]^2 \, .}
\label{chisg0_qn}
\end{equation}
In the paramagnetic phase, the single spin averages in Eq.~\eqref{chisg0_qn} vanish,
and $q = 0$ so the integral over $z$ gives unity.  We therefore have
\begin{equation}
\chiSG^0 =   \left[ \sum_l \Bigl( \Delta \tau \langle \,
S(0) S(l)\, \rangle_{\overline{H}} \,\Bigr) \right]^2 , \qquad (\text{for\ } q = 0).
\label{chisg0_qu_p}
\end{equation}
Equations \eqref{chisg_chisg0} and \eqref{chisg0_qu_p} have recently been
verified by comparing with series expansions~\cite{singh:17}.

\section{Conclusions}
\label{sec:concl}
We have derived in detail the RS solution of the classical and
quantum SK model. This solution is unstable at low temperatures, longitudinal
fields, and transverse fields, but the expressions derived here, in particular
the results for the replicon eigenvalue and the spin glass susceptibility, are
those use to \textit{show} the instability of the RS state. These notes are
largely an assembly of existing results; one result which seems to be new
is that for $\chiSG$ in the quantum case,
Eqs.~\eqref{chisg_chisg0} and \eqref{chisg0_qn}.

\bibliography{refs}

\begin{thebibliography}{19}
\expandafter\ifx\csname natexlab\endcsname\relax\def\natexlab#1{#1}\fi
\expandafter\ifx\csname bibnamefont\endcsname\relax
  \def\bibnamefont#1{#1}\fi
\expandafter\ifx\csname bibfnamefont\endcsname\relax
  \def\bibfnamefont#1{#1}\fi
\expandafter\ifx\csname citenamefont\endcsname\relax
  \def\citenamefont#1{#1}\fi
\expandafter\ifx\csname url\endcsname\relax
  \def\url#1{\texttt{#1}}\fi
\expandafter\ifx\csname urlprefix\endcsname\relax\def\urlprefix{URL }\fi
\providecommand{\bibinfo}[2]{#2}
\providecommand{\eprint}[2][]{\url{#2}}

\bibitem[{\citenamefont{Sherrington and Kirkpatrick}(1975)}]{sherrington:75}
\bibinfo{author}{\bibfnamefont{D.}~\bibnamefont{Sherrington}} \bibnamefont{and}
  \bibinfo{author}{\bibfnamefont{S.}~\bibnamefont{Kirkpatrick}},
  \emph{\bibinfo{title}{Solvable model of a spin glass}},
  \bibinfo{journal}{Phys. Rev. Lett.} \textbf{\bibinfo{volume}{35}},
  \bibinfo{pages}{1792} (\bibinfo{year}{1975}).

\bibitem[{\citenamefont{Binder and Young}(1986)}]{binder:86}
\bibinfo{author}{\bibfnamefont{K.}~\bibnamefont{Binder}} \bibnamefont{and}
  \bibinfo{author}{\bibfnamefont{A.~P.} \bibnamefont{Young}},
  \emph{\bibinfo{title}{Spin glasses: Experimental facts, theoretical concepts
  and open questions}}, \bibinfo{journal}{Rev. Mod. Phys.}
  \textbf{\bibinfo{volume}{58}}, \bibinfo{pages}{801} (\bibinfo{year}{1986}).

\bibitem[{\citenamefont{Parisi}(1980)}]{parisi:80}
\bibinfo{author}{\bibfnamefont{G.}~\bibnamefont{Parisi}},
  \emph{\bibinfo{title}{The order parameter for spin glasses: a function on the
  interval $0$--$1$}}, \bibinfo{journal}{J. Phys. A.}
  \textbf{\bibinfo{volume}{13}}, \bibinfo{pages}{1101} (\bibinfo{year}{1980}).

\bibitem[{\citenamefont{Parisi}(1983)}]{parisi:83}
\bibinfo{author}{\bibfnamefont{G.}~\bibnamefont{Parisi}},
  \emph{\bibinfo{title}{Order parameter for spin-glasses}},
  \bibinfo{journal}{Phys. Rev. Lett.} \textbf{\bibinfo{volume}{50}},
  \bibinfo{pages}{1946} (\bibinfo{year}{1983}).

\bibitem[{\citenamefont{de~Almeida and Thouless}(1978)}]{almeida:78}
\bibinfo{author}{\bibfnamefont{J.~R.~L.} \bibnamefont{de~Almeida}}
  \bibnamefont{and} \bibinfo{author}{\bibfnamefont{D.~J.}
  \bibnamefont{Thouless}}, \emph{\bibinfo{title}{Stability of the
  {S}herrington-{K}irkpatrick solution of a spin glass model}},
  \bibinfo{journal}{J. Phys. A} \textbf{\bibinfo{volume}{11}},
  \bibinfo{pages}{983} (\bibinfo{year}{1978}).

\bibitem[{\citenamefont{Edwards and Anderson}(1975)}]{edwards:75}
\bibinfo{author}{\bibfnamefont{S.~F.} \bibnamefont{Edwards}} \bibnamefont{and}
  \bibinfo{author}{\bibfnamefont{P.~W.} \bibnamefont{Anderson}},
  \emph{\bibinfo{title}{Theory of spin glasses}}, \bibinfo{journal}{J. Phys. F}
  \textbf{\bibinfo{volume}{5}}, \bibinfo{pages}{965} (\bibinfo{year}{1975}).

\bibitem[{\citenamefont{Sharma and Young}(2010)}]{sharma:10}
\bibinfo{author}{\bibfnamefont{A.}~\bibnamefont{Sharma}} \bibnamefont{and}
  \bibinfo{author}{\bibfnamefont{A.~P.} \bibnamefont{Young}},
  \emph{\bibinfo{title}{de {A}lmeida-{T}houless line in vector spin glasses}},
  \bibinfo{journal}{Phys. Rev. E} \textbf{\bibinfo{volume}{81}},
  \bibinfo{pages}{061115} (\bibinfo{year}{2010}), \eprint{(arXiv:1003.5599)}.

\bibitem[{\citenamefont{Singh and Young}(2017{\natexlab{a}})}]{singh:17b}
\bibinfo{author}{\bibfnamefont{R.~R.~P.} \bibnamefont{Singh}} \bibnamefont{and}
  \bibinfo{author}{\bibfnamefont{A.~P.} \bibnamefont{Young}},
  \emph{\bibinfo{title}{On the {A}lmeida-{T}houless instability in short-range
  {I}sing spin-glasses}} (\bibinfo{year}{2017}{\natexlab{a}}),
  \bibinfo{note}{(arXiv:1705.01164)}.

\bibitem[{\citenamefont{Yamamoto and Ishii}(1987)}]{yamamoto:87}
\bibinfo{author}{\bibfnamefont{T.}~\bibnamefont{Yamamoto}} \bibnamefont{and}
  \bibinfo{author}{\bibfnamefont{H.}~\bibnamefont{Ishii}},
  \emph{\bibinfo{title}{A perturbation expansion for the
  {S}herrington-{K}irkpatrick model with a transverse field}},
  \bibinfo{journal}{J. Phys. C} \textbf{\bibinfo{volume}{20}},
  \bibinfo{pages}{6053} (\bibinfo{year}{1987}).

\bibitem[{\citenamefont{Ray et~al.}(1989)\citenamefont{Ray, Chakrabarti, and
  Chakrabarti}}]{ray:89}
\bibinfo{author}{\bibfnamefont{P.}~\bibnamefont{Ray}},
  \bibinfo{author}{\bibfnamefont{B.~S.} \bibnamefont{Chakrabarti}},
  \bibnamefont{and}
  \bibinfo{author}{\bibfnamefont{A.}~\bibnamefont{Chakrabarti}},
  \emph{\bibinfo{title}{Sherrington-{K}irkpatrick model in a transverse field:
  {A}bsence of replica symmetry breaking due to quantum fluctuations}},
  \bibinfo{journal}{Phys. Rev. B} \textbf{\bibinfo{volume}{39}},
  \bibinfo{pages}{11828} (\bibinfo{year}{1989}).

\bibitem[{\citenamefont{Lai and Goldschmidt}(1990)}]{lai:90}
\bibinfo{author}{\bibfnamefont{P.-Y.} \bibnamefont{Lai}} \bibnamefont{and}
  \bibinfo{author}{\bibfnamefont{Y.~Y.} \bibnamefont{Goldschmidt}},
  \emph{\bibinfo{title}{Monte {C}arlo studies of the {I}sing spin-glass in a
  transverse field}}, \bibinfo{journal}{Europhys. Lett.}
  \textbf{\bibinfo{volume}{13}}, \bibinfo{pages}{289} (\bibinfo{year}{1990}).

\bibitem[{\citenamefont{Goldschmidt}(1990)}]{goldschmidt:90}
\bibinfo{author}{\bibfnamefont{Y.~Y.} \bibnamefont{Goldschmidt}},
  \emph{\bibinfo{title}{Solvable model of the quantum spin glass in a
  transverse field}}, \bibinfo{journal}{Phys. Rev. B}
  \textbf{\bibinfo{volume}{41}}, \bibinfo{pages}{4858} (\bibinfo{year}{1990}).

\bibitem[{\citenamefont{B\"uttner and Usadel}(1990)}]{buttner:90}
\bibinfo{author}{\bibfnamefont{G.}~\bibnamefont{B\"uttner}} \bibnamefont{and}
  \bibinfo{author}{\bibfnamefont{K.~D.} \bibnamefont{Usadel}},
  \emph{\bibinfo{title}{Stability analysis of an {I}sing spin glass with
  transverse field}}, \bibinfo{journal}{Phys. Rev. B}
  \textbf{\bibinfo{volume}{41}}, \bibinfo{pages}{428} (\bibinfo{year}{1990}).

\bibitem[{\citenamefont{Read et~al.}(1995)\citenamefont{Read, Sachdev, and
  Ye}}]{read:95}
\bibinfo{author}{\bibfnamefont{N.}~\bibnamefont{Read}},
  \bibinfo{author}{\bibfnamefont{S.}~\bibnamefont{Sachdev}}, \bibnamefont{and}
  \bibinfo{author}{\bibfnamefont{J.}~\bibnamefont{Ye}},
  \emph{\bibinfo{title}{Landau theory of quantum spin glasses of rotors and
  {I}sing spins}}, \bibinfo{journal}{Phys. Rev. B}
  \textbf{\bibinfo{volume}{52}}, \bibinfo{pages}{384} (\bibinfo{year}{1995}).

\bibitem[{\citenamefont{Federov and Shender}(1986)}]{federov:86}
\bibinfo{author}{\bibfnamefont{Y.}~\bibnamefont{Federov}} \bibnamefont{and}
  \bibinfo{author}{\bibfnamefont{E.}~\bibnamefont{Shender}},
  \emph{\bibinfo{title}{Quantum spin glasses in the {I}sing model with a
  transverse field}}, \bibinfo{journal}{JETP Lett.}
  \textbf{\bibinfo{volume}{43}}, \bibinfo{pages}{681} (\bibinfo{year}{1986}).

\bibitem[{\citenamefont{Ye et~al.}(1993)\citenamefont{Ye, Sachdev, and
  Read}}]{ye:93}
\bibinfo{author}{\bibfnamefont{J.}~\bibnamefont{Ye}},
  \bibinfo{author}{\bibfnamefont{S.}~\bibnamefont{Sachdev}}, \bibnamefont{and}
  \bibinfo{author}{\bibfnamefont{N.}~\bibnamefont{Read}},
  \emph{\bibinfo{title}{Solvable spin glass of quantum rotors}},
  \bibinfo{journal}{Phys. Rev. Lett.} \textbf{\bibinfo{volume}{70}},
  \bibinfo{pages}{4011} (\bibinfo{year}{1993}).

\bibitem[{\citenamefont{Huse and Miller}(1993)}]{huse:93}
\bibinfo{author}{\bibfnamefont{D.~A.} \bibnamefont{Huse}} \bibnamefont{and}
  \bibinfo{author}{\bibfnamefont{J.}~\bibnamefont{Miller}},
  \emph{\bibinfo{title}{Zero-temperature critical behavior of the
  infinite-range quantum {I}sing spin glass}}, \bibinfo{journal}{Phys. Rev.
  Lett.} \textbf{\bibinfo{volume}{70}}, \bibinfo{pages}{3147}
  (\bibinfo{year}{1993}).

\bibitem[{\citenamefont{Sachdev}(1999)}]{sachdev:99}
\bibinfo{author}{\bibfnamefont{S.}~\bibnamefont{Sachdev}},
  \emph{\bibinfo{title}{Quantum Phase Transitions}}
  (\bibinfo{publisher}{Cambridge University Press},
  \bibinfo{address}{Cambridge}, \bibinfo{year}{1999}).

\bibitem[{\citenamefont{Singh and Young}(2017{\natexlab{b}})}]{singh:17}
\bibinfo{author}{\bibfnamefont{R.~R.~P.} \bibnamefont{Singh}} \bibnamefont{and}
  \bibinfo{author}{\bibfnamefont{A.~P.} \bibnamefont{Young}},
  \emph{\bibinfo{title}{Critical and {G}riffiths-{M}c{C}oy singularities in
  quantum {I}sing spin-glasses on d-dimensional hypercubic lattices: {A} series
  expansion study}} (\bibinfo{year}{2017}{\natexlab{b}}),
  \bibinfo{note}{(unpublished)}.

\end{thebibliography}

\end{document}